\documentclass[aps,twocolumn,showpacs,prl]{revtex4}
\usepackage[english]{babel}
\usepackage{amsmath}
\usepackage{amssymb} 
\usepackage{times}
\usepackage{epsfig} 

\begin{document} 

\title{Ultrafast charge recombination in photoexcited Mott-Hubbard insulator}

\author{Zala Lenar\v ci\v c$^{1}$ and Peter Prelov\v sek$^{1,2}$}
\affiliation{$^1$J.\ Stefan Institute, SI-1000 Ljubljana, Slovenia}
\affiliation{$^2$Faculty of Mathematics and Physics, University of
Ljubljana, SI-1000 Ljubljana, Slovenia}

\begin{abstract}
{We present a calculation of the 
recombination rate of the excited holon-doublon pairs 
based on the two-dimensional model relevant for undoped cuprates which shows that fast
processes, observed in pump-probe experiments on Mott-Hubbard insulators in picosecond range, 
can be explained even quantitatively with the 
multi-magnon emission.
The precondition is the existence of the Mott-Hubbard bound exciton
of the s-type. We find that its decay is exponentially dependent on the 
Mott-Hubbard gap and on the magnon energy, with a small prefactor
which can be traced back to strong correlations and consequently
large exciton-magnon coupling.}
\end{abstract}

\pacs{71.27.+a, 78.47.J-, 74.72.Cj}

\maketitle
The advance in the ultrafast absorption spectroscopy allow for a direct access
to excited states in a variety of materials as well as detailed studies of
relaxation and thermalization processes. Correlated systems and Mott-Hubbard 
(MH) insulators
are in this connection particularly of interest since here phenomena differ
qualitatively from those well understood in band insulators and 
semiconductors.
A prototype MH system are undoped cuprates  as 
YBa$_2$Cu$_3$O$_6$, La$_2$CuO$_4$ and Nd$_2$CuO$_4$  
which have been investigated with femtosecond pump-probe 
spectroscopy \cite{matsuda94,okamoto10,okamoto11} with an universal observation 
of ultrafast relaxation and recombination. The photoinduced
carriers in undoped cuprates excited across the MH gap, which is in these systems 
of charge-transfer type, recombine in a picosecond range being a scale 
far below the usual physics of clean semiconductors with similar band gaps
\cite{yu99}.
The origin of such ultrafast processes has been already qualitatively
attributed to  strong-correlation effects, in particular to larger energy of 
magnon excitations
with the exchange energy $J$ and strong intrinsic coupling of
charge carriers with spin  fluctuations within the MH insulator \cite{imada98}, still no
model-based theory was presented so far. Other MH insulators probed recently 
by ultrafast spectroscopy \cite{wall11,novelli12} also exhibit fast-relaxation  
phenomena.   Analogous are findings in the experiments with ultracold fermions 
in optical lattices where the decay of double occupancies \cite{strohmaier10,
sensarma10} are related to the correlated nature of the background state.

Due to strong Coulomb repulsion (strong correlations) excited quasiparticles
within the MH insulators can be considered as empty sites (holons) and doubly
occupied sites (doublons), respectively. 
In the following we consider  as a prototype model 
the single-band Hubbard model on two-dimensional (2D) square 
lattice in the regime  of large $U \gg t$, whereby the generalization to more 
appropriate charge-transfer model for cuprates is quite straightforward
and the comparison with the parameters for actual materials can be also performed.
 
\begin{figure}[ht]
\includegraphics[width=0.4\textwidth]{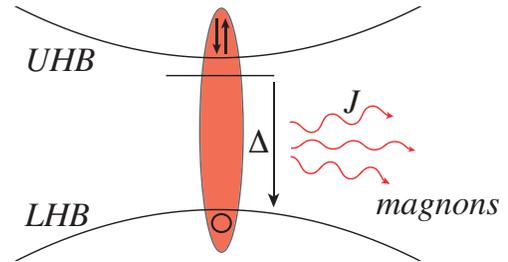}
\caption{(Color online) The process of the Mott-Hubbard exciton recombination
via the multi-magnon emission.}
\label{fig0}
\end{figure}

When interpreting the pump-probe experiments it is assumed that 
initially photoinduced mobile charges - holons and doublons - 
after very fast transient in the femtosecond range form
a MH exciton, i.e., a bound holon-doublon (HD) pair. 
The existence and stability of the MH exciton has been previously addressed
in connection with cuprates both in model studies 
\cite{wrobel02,tohyama02,tohyama06}
as well in the interpretation of optical response \cite{choi99,novelli12} 
and of large Raman shift \cite{salamon95} with some conflicting conclusions. 
As a result of our study the disagreement can be reconciled with the observation that the
ground state (g.s.) of bound HD pair is of the s-type, thus not observable
in an optical absorption.  Just such a s-type pair state appears to be the
precondition for a fast non radiative recombination process. The evidence
for the intermediate HD bound state can be 
pump-intensity independent recombination rates, apparently observed in 
experiments \cite{okamoto11}. Nevertheless, the formation of an MH exciton is 
a nontrivial consequence of strong correlations and large 
charge-magnon coupling in 2D antiferromagnetic (AFM) background. The recombination 
of an MH exciton, i.e., a bound pair of a doublon in the upper Hubbard band 
(UHB) and a holon in a lower Hubbard band (LHB), across the MH gap $\Delta$ 
via the the multi-magnon emission is schematically presented in Fig.~\ref{fig0}.
Using the golden-rule approach to the recombination rate $\Gamma = 1/\tau$ 
of the exciton (numericallly evaluated on small 
systems) we show that it is  very large 
in comparison with clean semiconductors. In particular our results are consistent 
with the exponential dependence,
\begin{equation}
\Gamma \propto \exp(-\alpha \Delta/J), \label{fit1}
\end{equation}
so that enhanced $\Gamma$ has origin in large exchange scale $J$, but 
even more in small $\alpha  \sim 0.5$ which can be again traced back to 
strong correlations. In spite of  large $\Delta$ 
recombination times $\tau$ are in the picosecond range, 
consistent with pump-probe experiments in undoped cuprates 
\cite{okamoto10,okamoto11}.   

As the starting model for the study of photoexcited states  in the MH insulator 
we use the single-band Hubbard model, 
\begin{equation}
H= -  t \sum_{ \langle ij \rangle s} c^\dagger_{j s}  c_{i s}   
 + U \sum_i n_{i \uparrow} n_{i \downarrow}, \label{hub}
\end{equation}
where sums $\langle ij\rangle$ run over nearest-neighbor (n.n.) pairs of sites
on a square 2D lattice.
More realistic model for cuprates beyond the Hubbard model we comment furtheron.
To reduce the complexity and as well to get correlation effects more transparent  
in the regime $U\gg t$ relevant for the MH insulator we 
perform the standard canonical transformation \cite{chao77}.
Within the lowest order in $t^2/U$ and neglecting the 3-site terms
one gets a generalized $t$-$J$ model for an insulator, including besides 
a spin-exchange term also both holon and doublon operators,
\begin{eqnarray}
H_{tJ}&=& t \sum_{ \langle ij \rangle s} (h^\dagger_{j s}  h_{i s}   -
d^\dagger_{j s}  d_{i s} + {\rm h.c.}) + U \sum_i n_{di}  + \nonumber\\
&& + J \sum_{\langle ij \rangle} ({\bf S}_j \cdot {\bf S}_i - \frac{1}{4}\delta_{1,n_j n_i} ), \label{htj}
 \end{eqnarray}   
where $h_{is}^\dagger = c_{is}(1-n_{i\bar s})$ and 
$d^\dagger_{is} = c^\dagger_{i\bar s} n_{is}$
are projected fermionic operators, excluding double and unoccupied sites, respectively.
Note that $h^\dagger _{is}, d^\dagger_{is}$ up to a sign represent 
creation holon and doublon 
operators on a site previously occupied by spin $s$ while
$n_{di}= \sum_s d^\dagger_{is}d_{is}$ and $n_{hi}= \sum_s h^\dagger_{is}h_{is}$ 
are the local number of doublons and holons, respectively.
It is evident that $H_{tJ}$ conserves the number of holons $N_h$ and doublons
$N_d$. For the study of the HD recombination it is essential 
to include the 3-site term, which emerges after the canonical transformation \cite{chao77}
also within the order $t^2/U$. It has been so far mostly derived and considered
for the hole-doped reference insulator \cite{hirsch85,ramsak89} and can be for the
HD case (neglecting terms which do not represent HD recombination) 
written in a spin-invariant form, 
\begin{equation}
H_3= t_3 \sum_{(ijk) ss'} [h_{i s } d_{k s'}   
{\vec \sigma}_{s\bar{s}'} \cdot {\bf S}_j  + {\rm h.c.} ],  \label{h3}
\end{equation}
where $i\neq k$ are n.n. sites to $j$ and ${\vec \sigma}$ the Pauli matrix.
Within the lowest order of the $t/U$ expansion one gets  
$t_3=2t^2/U=J/2$. Evidently, $H_3$ creates and annihilates
HD pairs which are on next-nearest-neighbor (n.n.n.)
sites. It should be stressed that within a canonically transformed Hamiltonian
there is no recombination term being on n.n. sites or being linear in $t$.
We use furtheron the hopping as the unit of energy, $t=1$.

We proceed to the evaluation of the HD recombination process in two steps.
To describe the relaxation of the photoexcited insulator through
the intermediate bottleneck stage of an MH exciton  we 
first establish the stability of  the bound HD pair within the $H_{tJ}$,
Eq.~(\ref{htj}), neglecting $H_3$. The problem is analogous to 
extensively studied case of the binding of two holes $N_h=2$ 
within the $t$-$J$ model
\cite{dagotto94}, with an important distinction that holon and doublon
are distinguishable particles and can form also $s$-type (A$_1$
symmetry) bound pair. In the $N_h=2$ problem the g.s. has the $d$-type
bound state above the threshold $J>J_c \sim 0.3 $ \cite{dagotto94}.

\begin{figure}[ht]
\includegraphics[width=0.43\textwidth]{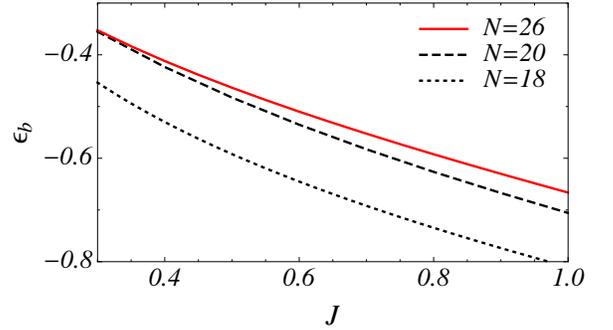}
\caption{(Color online) Holon-doublon pair binding energy $\epsilon_b$ vs. $J$.
Plotted are results for the s-type g.s. for $N=18,20,26$ lattices.}
\label{fig1}
\end{figure}

We study eigenstates of a single HD pair, i.e., $N_h=N_d=1$ 
by performing the exact diagonalization of $H_{tJ}$ on 
small square and rectangular lattices with periodic boundary conditions 
using the Lanczos technique
on $N \leq 26$ sites. As a criterion for binding we use the 
pair binding energy $\epsilon_b= E^{hd}_0 -E^h_0 -E^d_0+ E^0_0$
where $E^{hd}_{0},E^h_0, E^d_0, E^0_0$ refer to the HD pair, single hole, single doublon 
and undoped AFM g. s., respectively. Within $H_{tJ}$, Eq.~(\ref{htj}),
we also take into account $E^d_0=E^h_0+ U$, while for $\epsilon^h_0=E^h_0-E^0_0$ 
we use known accurate fits \cite{leung95}.
Results for $\epsilon_b(J)$ obtained for $N=18, 20, 26$ 
are presented in Fig.~\ref{fig1}  for the lowest s-type state. Values for $\epsilon_b$ 
appear to be quite size independent and confirm previous findings
\cite{tohyama06} of the stability $\epsilon_b<0$ of the MH exciton. The state is, 
however, of an even symmetry and not observable in an
optical transition from the insulator  AFM state. On the other hand, the p-type 
state relevant for optical absorption reveals $\epsilon_b  \gtrsim 0$, hence 
does not appear to be a bound one. 

\begin{figure}[ht]
\includegraphics[width=0.2\textwidth]{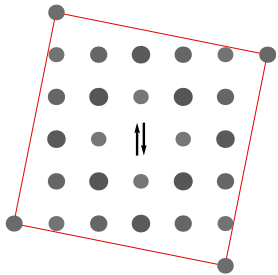}
\caption{(Color online) Density correlations $D_j$ of the hole  with respect to the doublon
position for $J=0.4$, as calculated for $N=26$ lattice.}
\label{fig2}
\end{figure}

Another test for the HD binding are g.s. density correlations
$D_j=\langle \psi^{hd}_0| n_{hj} n_{d0}  | \psi^{hd}_0 \rangle$
(the position of the doublon chosen as the origin).  In Fig.~\ref{fig2} $D_j$ 
obtained (and symmetrized to recover the reflection symmetry) on $N=26$ 
for $J=0.4$ are consistent with the binding where the largest probability is for the 
HD pair being on a distance $d_0 = \sqrt{2}$. On contrary, the unbound states 
(as the p-type lowest state) are characterized by largest $d_0 \sim \sqrt{N}$) 
for a given lattice.

The next stage is to evaluate the recombination rate $\Gamma= \tau^{-1}$ of the MH exciton
using the Fermi golden rule with $H_3$, Eq.~(\ref{h3}), serving as the perturbation, 
\begin{equation}
\Gamma=2\pi \sum_{m}|\langle\psi_{m}^0|H_3|\psi_{0}^{hd}\rangle|^2 \ 
\delta(E_{m}^{0}-E_{0}^{hd}), \label{gamma}
\end{equation}
and transitions are into magnon states $|\psi_{m}^0\rangle$ within the undoped AFM.
We note that Eq.~(\ref{gamma}) can be represented as a resolvent $\Gamma= -2~ {\rm Im}
C(\omega=E_0^{hd}-E_0^0)$,
\begin{equation}
C(\omega)=\langle\psi_0^{hd}|H_3 \frac{1}{\omega^+ +E_0^0 -H_{J} }H_3|
\psi_0^{hd}\rangle, \label{com}
\end{equation}
with $\omega^+=\omega+i \delta$. In the calculation only the AFM part
$H_J$ of the  $H_{tJ}$, Eq.~({\ref{htj}),  is relevant. 
Eq.~(\ref{com}) is convenient for the evaluation within the Lanczos procedure,
as usual for g.s. correlation functions \cite{dagotto94,prelovsek11}.  Starting with 
the perturbed $|\phi_0\rangle=H_3|\psi_0^{hd}\rangle $ the Lanczos procedure generates
a tridiagonal matrix and $C(\omega)$ can be expressed in a continued-fraction
form.
While $\delta$ should be infinitesimally small thermodynamic
limit $N \to \infty$  can be efficiently simulated by adopting finite $\delta \ll J$. Moreover, to avoid 
the influence of such a Lorentzian  smoothing (note that final $\Gamma \ll 1$) 
we employ rather a Gaussian smoothing 
of poles produced after a finite number of Lanczos steps, $M \sim 150$ .

The convergence of results with the system size is presented in Fig.~\ref{fig3}
where $\Gamma$ is evaluated for fixed $J=0.4$ and different lattices with 
$N=20,24,26$ ($N=24=4\times 6$ sites  is a rectangular lattice) sites. 
Results are shown versus the effective MH gap 
$\Delta = E_0^{hd}-E_0^0 $ between the bound HD-pair state and the AFM g.s. 
(see Fig.~\ref{fig0}). Taking $\Delta$ as a variable corresponds to assume $U$ in Eq.~(\ref{htj})
as an independent  parameter  which could deviate from the relation $U=4t^2/J$
obtained within the $U\gg t $ limit. 
We note from Fig.~\ref{fig3}
that even small $\delta = 0.07$ is enough
to obtain continuous $\Gamma(\Delta)$ for the largest system $N=26$.  Clearly,
the system size $N$ should not be too small in order to accommodate enough magnons 
and to get reliable results in the relevant tails $\Gamma \ll 1$. 

\begin{figure}[ht]
\includegraphics[width=0.43\textwidth]{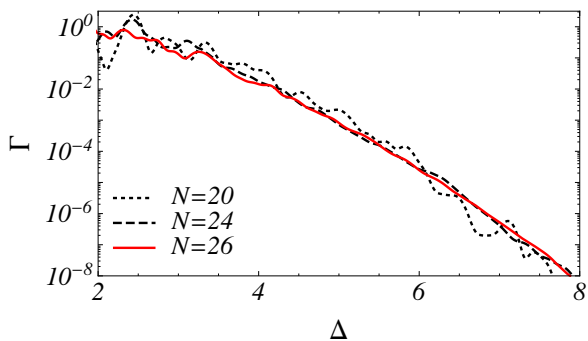}
\caption{(Color online) Exciton recombination rate $\Gamma$ vs. gap $\Delta$
for $J=0.4$ and different planar systems with $N=20, 24, 26$ sites. 
Results are smoothed with $\delta = 0.07$.}
\label{fig3}
\end{figure}

As the central result we present in Fig.~\ref{fig4}a,b  calculated $\Gamma$ vs. $\Delta$ 
obtained on the largest system $N=26$ and different $J=0.3,0.4, 0.6$. 
The dependence on $\Delta/J$ in the relevant
regime $\Gamma(\Delta \sim U) $ is close to the exponential, Eq.~(\ref{fit1}),
as presented in Fig.~\ref{fig4}a. Effective $\alpha$ in this case is within the
window $ 0.3 <\alpha <0.7$. Even better fit can be reached by 
following the perturbation-theory estimate for the probability of 
the generation of $n_0$ bosons
\cite{strohmaier10,sensarma10,avouris06} where in our case $n_{0} =\Delta/J$,
\begin{equation}
\Gamma \propto \left[\frac{1}{n_{0} !}\left(\frac{t}{J}\right)^{n_{0}}\right]^{2} \propto \left(\frac{et}{n_{0}J} \right)^{2n_{0}} \propto 
{\rm exp}\left[ - \alpha_0 \frac{\Delta}{J} \ln{\frac{\Delta}{e t} }\right] . \label{fit2}
\end{equation} 
Results presented in Fig.~\ref{fig4}b for $\Gamma$ vs. $(\Delta/J)  {\rm ln}(\Delta/te)$
reveal even better agreement, with quite universal $\alpha_0 \sim 0.8 $ (instead of $\alpha_{0}=2$ 
following from straight derivation of Eq.~(\ref{fit2})~).

\begin{figure}[ht]
\includegraphics[width=0.43\textwidth]{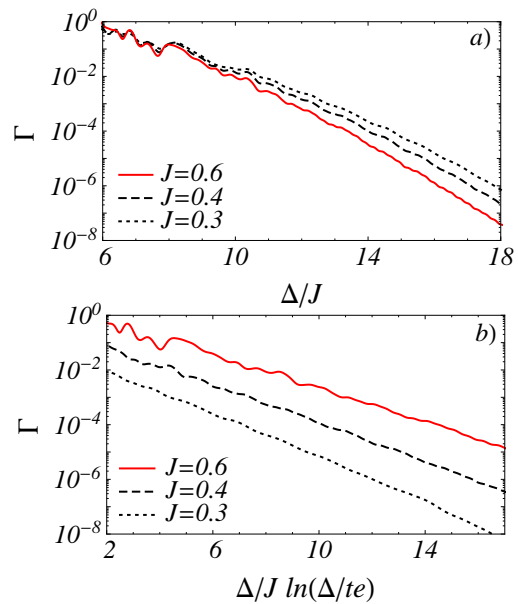}
\caption{(Color online) Exciton recombination rate $\Gamma$ vs. 
a) $\Delta/J$ and b)  $\Delta/J\ln(\Delta/te)$
for different $J=0.3,0.4,0.6$ as calculated for $N=26$ sites.}
\label{fig4}
\end{figure}

Exponential dependence of $\Gamma$ on $J$, Eq.~(\ref{fit1}),
is plausible since the recombination involves the multi-magnon emission.
Taken at the face value within the reduced  model, Eq.~(\ref{htj}), 
with $\Delta = U  - 2 |\epsilon^h_0| - |\epsilon_b| $  and $U= 4 t^2/J $
the variation of $\Gamma$ with $J$ is even enhanced. Nevertheless,
even for quite realistic value $J=0.4$ we get in this simplified case 
$\Delta /J \sim 17$ and  $\Gamma \sim 10^{-6}$ which is not
unreasonably small taking into account large number of involved magnons. 
On the other hand,
any decrease of the 'effective' gap $\Delta$ leads to
a large enhancement of the rate $\Gamma$.

Before the application of our theory and results to actual 
MH-insulator materials and experiments we offer some explanation of 
substantial $\Gamma$ even for cases where evidently a large
number of emitted magnons $n_{0} \sim \Delta/J \gg1$ is required for
the HD-pair recombination.  It is already evident
from numerous numerical \cite{dagotto94} and analytical \cite{horsch91} 
studies of the single hole $N_h=1$ within the $t$-$J$
model  that a holon (or doublon) perturbs strongly the AFM background in the
strong correlation regime  $J < t$. Even more this is the case for two (bound) holes 
$N_h=2$ \cite{dagotto94} and for the present example of a bound HD pair, $N_h=N_d=1$. 
In Fig.~\ref{fig5} we present results for exchange-energy deviations relative 
to the reference AFM g.s. Results are for $J=0.4$ and for the most probable 
HD configuration with the pair distance $d_0=\sqrt{2}$. It is quite evident that 
the recombination of a HD pair via $H_3$ (which just requires 
the holon and doublon being  on n.n.n. sites) already leads to the generation 
of a large number of magnons $n_0 \gg 1$.    
\begin{figure}[ht]
\includegraphics[width=0.3\textwidth]{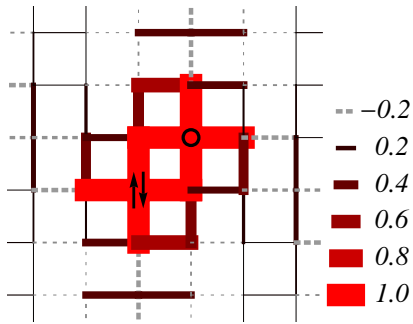}
\caption{(Color online) Bond perturbations within the 
holon-doublon pair g.s.,  presented for the most probable pair 
configuration. Results are for $J=0.4$ and deviations of bond energies
relative to AFM g.s. are presented.}
\label{fig5}
\end{figure}

Related but not equivalent argument for the appreciable $\Gamma$ for
cases $\Delta \gg J$ is small  parameter $\alpha_0 <1$ in the Ansatz (\ref{fit2})
and even more $\alpha \sim 0.5$ (within the relevant regime of $J/t$) in Eq.~(\ref{fit1}).  
This can be considered as another manifestation of the strong coupling 
between the charge and spin excitations within the MH insulator as well as 
in  strongly correlated systems in general.

{\it Application to cuprates:}  Presented analysis and results can be generalized 
in quite straightforward way to insulators, where the MH gap represents 
charge-transfer excitations. In particular this includes insulating and superconducting
cuprates,
where more complete (three-band) model includes hybridized Cu d orbitals and O p orbitals.
Although this goes beyond the prototype single-band Hubbard model in our study,
through the Zhang-Rice mechanism \cite{zhang88} quasiparticles  can be well
described within the $t$-$J$ model both for doped holes  and doped 
electrons \cite{tohyama04}. Parameters are quite well established 
as $t\sim 0.35$eV and $J/t \sim 0.4$ \cite{tohyama04}. The three site term $H_3$, Eq.~(\ref{h3}),
as well emerges from the three-band model \cite{ramsak89}, although 
$t_3$ can be quantitatively different but also less important entering $\Gamma$ in 
Eq.~(\ref{gamma}) only as a prefactor. So the essential generalization emerges 
from fact that the MH gap $\Delta$ and the 'effective bare' $U$ in the model, Eq.~(\ref{htj}), 
are not directly related 
to $J$ and can be considered to large extent as independent parameters, taken
from experiment or from more complete model. Optical MH gaps in undoped cuprates 
are experimentally  known $\Delta_0 \sim 1.5$eV, still  differing between
members of cuprate family. In order to get defined $\Delta$ in Eq.~(\ref{fit1})  and in Fig.~\ref{fig0}
we have to take into account that optical transitions are to HD p-type unbound states, 
so that $\Delta = \Delta_0 - |\epsilon_b|$. 

We first consider Nd$_{2}$CuO$_{4}$ with the optical gap $\Delta_{0}=1.6~$eV \cite{okamoto11}
and $J = 0.155$~eV,  so that we get $\Delta=4.1~t$ and $\Gamma  \sim 2.2 \cdot 10^{-2}$ in units of 
$\tau_{0}^{-1}$ where $\tau_0= \hbar/t \sim 2$~fs. This yields final relaxation time 
$\tau=\Gamma^{-1} \sim 0.09~$ps which is surprisingly close to experimentally established 
$\tau \sim 0.2$~ps \cite{okamoto11}. Analogous, for La$_{2}$CuO$_{4}$ with 
$\Delta_{0}=2$~eV and $J=0.133$~eV we get $\Delta=5.3~t$ and $\Gamma\sim 1.3 \cdot 10^{-4}$ 
which yields $\tau\sim 15$~ps. For the latter compound $\tau$ has not been well determined
\cite{okamoto11} but is considerably longer than in Nd$_{2}$CuO$_{4}$ consistent with our result.

{\it Conclusions and discussion:}
The most important finding of this study is that multi-magnon emission can be quite
an efficient mechanism for the nonradiative recombination of photoinduced charges in a MH insulator.
This is in contrast to usual band insulators and semiconductors where the importance 
of analogous multi-phonon processes has not been 
established theoretically or experimentally \cite{yu99}, while such processes could be 
relevant for some novel structures as carbon nanotubes \cite{avouris06,perebeinos08}.  
The distinction of MH insulators  is not just in larger boson scale $J$  compared to phonon 
energies but primarily in the strong coupling
between charge quasiparticles (holes and doublons) and spin excitations. The main 
manifestation is in the established prefactor $\alpha_0<1$ in Eq.~(\ref{fit2}) as well as in
$\alpha \sim 0.5$  in the exponential Eq.~(\ref{fit1}). 
It should be also reminded that in spite
of strong correlations doped holes and doublons are quite mobile with 
modestly enhanced effective mass. 

When interpreting recent pump-probe experiments on MH insulators, it should be 
reminded that  we consider here only the bottleneck process of
a MH-exciton recombination. The initial relaxation of the pump-induced transient metallic-like 
collective state of holons and doublons \cite{okamoto11} is expected to be much faster 
and  has been already investigated although not fully settled in different model studies 
\cite{takahashi02,hassanieh08,kanamori10,eckstein11,eckstein12}.

It should be noted that a similar mechanism of the HD-pair recombination
has been recently considered as a candidate for the relaxation in fermionic
cold-atom systems \cite{strohmaier10,sensarma10} although it appears
to be subdominant process to the kinetic-energy assisted decay which requires 
a finite density (metallic-like state) of nonequilibrium
holons and doublons. The latter could emerge at  effective $T>0$
or in systems far from equilibrium. On contrary, for 2D cuprates our study reveals 
that the multi-magnon mechanism of charge recombination and thermalization 
is very efficient and leads to ultrafast relaxation in the picosecond range as 
experimentally observed. 
   
The authors acknowledge fruitful discussions with T. Tohyama.
This work has been supported by the Program P1-0044 and the project J1-4244 
of the Slovenian  Research Agency (ARRS).


\begin{thebibliography}{99} 
\bibitem{matsuda94} K. Matsuda et al., Phys. Rev. B \textbf{50}, 4097 (1994).
\bibitem{okamoto10} H. Okamoto et al., Phys. Rev. B \textbf{82}, 060513 (2010)
\bibitem{okamoto11} H. Okamoto et al., Phys. Rev. B \textbf{83}, 125102 (2011)
\bibitem{yu99} P. Yu and M. Cardona, \textit{Fundamentals of semiconductors: physics and materials properties} (Springer, Berlin, 1996)
\bibitem{imada98} M. Imada, A. Fujimori and Y. Tokura, Nature Phys. \textbf{7}, 114 (2011).
\bibitem{wall11} S. Wall, Nature Phys. \textbf{7}, 114 (2011).
\bibitem{novelli12} F. Novelli et al., arXiv:1205.4609 (2012).
\bibitem{strohmaier10} N. Strohmaier et al., Phys. Rev. Lett. \textbf{104}, 080401 (2010).
\bibitem{sensarma10} R. Sensarma et al., Phys. Rev. B \textbf{82}, 224302 (2010).
\bibitem{wrobel02} P. Wr\'obel and R. Eder, Phys. Rev. B \textbf{66}, 035111 (2002).
\bibitem{tohyama02} T. Tohyama, H. Onodera, K. Tsutsui and S. Maekawa, Phys. Rev. Lett. \textbf{89}, 257405 (2002).
\bibitem{tohyama06} T. Tohyama, J. Phys. Soc. Jpn. \textbf{75}, 34713 (2006).
\bibitem{choi99} H. S. Choi et al., Phys. Rev. B \textbf{60}, 4646 (1999).
\bibitem{salamon95} D. Salamon et al., Phys. Rev. B \textbf{51}, 6617 (1995).
\bibitem{chao77} K. A. Chao, J. Spa\l{}ek and  A. M. Ole\'s, J. Phys. C \textbf{10}, L271 (1977).
\bibitem{hirsch85} J. E. Hirsch, Phys. Rev. Lett. \textbf{54}, 1317 (1985).
\bibitem{ramsak89}A. Ram\v sak and P. Prelov\v sek, Phys. Rev. B \textbf{40}, 2239 (1989).
\bibitem{dagotto94} E. Dagotto, Rev. Mod. Phys. \textbf{66}, 763-840 (1994).
\bibitem{leung95} P. W. Leung and R. J. Gooding, Phys. Rev. B \textbf{52}, R15711 (1995).
\bibitem{prelovsek11} P. Prelov\v sek and J. Bon\v ca, arXiv:1111.5931 (2011).
\bibitem{avouris06} P. Avouris, M. Freitag and V. Perebeinos, Phys. Stat. Sol. \textbf{243}, 3197 (2006).
\bibitem{horsch91} G. Martinez and P. Horsch, Phys. Rev. B \textbf{44}, 317-331 (1991).
\bibitem{zhang88} F. C. Zhang and T. M. Rice, Phys. Rev. B \textbf{37}, 3759 (1988).
\bibitem{tohyama04} T. Tohyama, Phys. Rev. B \textbf{70}, 174517 (2004).
\bibitem{perebeinos08} V. Perebeinos and P. Avouris, Phys. Rev. Lett. \textbf{101}, 057401 (2008).
\bibitem{takahashi02} A. Takahashi, H. Gomi and M. Aihara, Phys. Rev. Lett. \textbf{89}, 206402 (2002).
\bibitem{hassanieh08} K. A. Al-Hassanieh, F. A. Reboredo, A. E. Feiguin, I. Gonz\'alez and E. Dagotto, Phys. Rev. Lett. \textbf{100}, 166403 (2008).
\bibitem{kanamori10} Y. Kanamori, H. Matsueda and S. Ishihara, Phys. Rev. B \textbf{82}, 115101 (2010).
\bibitem{eckstein11} M. Eckstein and P. Werner, Phys. Rev. B \textbf{84}, 035122 (2011).
\bibitem{eckstein12} M. Eckstein and P. Werner, arXiv:1207.0402 (2012).
\end{thebibliography}
\end{document}